\documentstyle[12pt,twoside,fleqn,espcrc1,epsfig]{article}

\def\bE{{\bf E}}
\def\bH{{\bf H}}
\def\bj{{\bf j}}

\def\bv{{\bf v}}

\newcommand{\AmS}{{\protect\the\textfont2
  A\kern-.1667em\lower.5ex\hbox{M}\kern-.125emS}}

\title{Flux tube dynamics in the dual superconductor}

\author{Melissa Lampert$^{\rm a*}$
and Benjamin Svetitsky\address{School of Physics and Astronomy,
Raymond and Beverly Sackler Faculty of Exact Sciences,
Tel Aviv University, 69978 Tel Aviv, Israel}\thanks{Work supported 
by the Israel Science Foundation 
under Grant No.~255/96-1 and by the Basic Research Fund of Tel Aviv 
University.\hfil\break
Talk given at Quark Matter '99, the 14th International Conference on
Ultra-Relativistic Nucleus--Nucleus Collisions, Torino, Italy, May
10--14, 1999.}
}

\begin{document}
\maketitle

\begin{abstract}
We have studied
plasma oscillations in a flux tube created in a dual superconductor.
Starting from a static flux tube configuration, with electric charges
at either end, we release a fluid of electric charges in the system
that accelerate and screen the electric field.
The weakening of the electric field allows the flux tube to collapse,
and the inertia of the charges forces it open again.
We find strong radiation of electric flux into the superconductor
in all regimes of couplings.
\end{abstract}

\section{INTRODUCTION}

't Hooft \cite{Hooft1}
and Mandelstam \cite{Mandelstam}
proposed long ago that quark confinement in QCD
would come about as the result of the confinement of color electric flux
into flux tubes, and that such flux tubes would form in a dual superconductor.
One possible pathway to formation of this dual superconductor was offered
later by 't~Hooft \cite{Hooft2}.
He showed that an Abelian projection of a non-Abelian gauge theory contains
magnetic monopoles.
Though the effective interaction among these monopoles is hard to calculate,
it is not unreasonable to suppose that they form a condensate like that of
the Cooper pairs in a superconductor.
This {\em magnetic} condensate would then bring about an {\em electric}
Meissner effect, confining electric flux and electric charge.

In order to study particle dynamics in QCD, one has to study dynamics of the
flux tube.
Casher, Kogut, and Susskind \cite{KS} argued that particle creation in the flux tube
is the soft process responsible for the inside--outside cascade
in $e^+e^-$ annihilation;
Casher, Neuberger, and Nussinov \cite{CNN}
calculated the particle spectrum via WKB
(generalizing Schwinger's \cite{Schwinger}
famous formula), and this picture then entered
phenomenology via the Lund Monte Carlo program \cite{Lund} and its descendants.
The flux tube has also been much studied in the context of $pA$ \cite{KMS}
and $AA$ \cite{BNK,CEKMS} collisions.
These and subsequent studies generally lacked any dynamics for the structure
of the flux tube itself.
We have taken the first step of studying the dynamics
of classical charges moving in an electric flux tube and the reaction
of the flux tube in the dual superconductor \cite{LS}.
\section{THE DUAL SUPERCONDUCTOR}

To specify the model, we begin with
Maxwell's equations coupled to both magnetic and electric currents,
\begin{eqnarray}
   \partial_\mu F^{\mu\nu} &=& j_e^\nu \label{egauss4}\\
   \partial_\mu \tilde F^{\mu\nu} &=& j_g^\nu \label{mgauss4}\ .
\end{eqnarray}
Eq.~(\ref{mgauss4}) is no longer just a Bianchi identity; thus
a vector potential can be introduced only if a new term is
added to take care of the magnetic current, {\em viz.,}
\begin{equation}
F^{\mu\nu} = \partial^\mu A^\nu - \partial^\nu A^\mu 
      + \epsilon^{\mu\nu\lambda\sigma} G_{\lambda\sigma}\ ,
\qquad
   G^{\mu\nu} = - n^\mu (n \cdot \partial)^{-1} j_g^\nu\ .
\end{equation}
This vector potential can be coupled to electric charges as usual;
in order to introduce {\em magnetic\/} charges, one introduces a
dual potential \cite{Zwanziger} via
\begin{equation}
   \tilde F^{\mu\nu} = \partial^\mu B^\nu - \partial^\nu B^\mu  
      + \epsilon^{\mu\nu\lambda\sigma} M_{\lambda\sigma}\ ,
\qquad
   M^{\mu\nu} = - n^\mu (n \cdot \partial)^{-1} j_e^\nu\ .
\end{equation}
Now we can write a model for the monopoles, for which the simplest is an
Abelian Higgs theory \cite{Suzuki,Osaka},
\begin{equation}
   D_\mu^B D^{\mu B}\psi + \lambda(|\psi|^2-v^2)\psi = 0 \label{Higgs}\ ,
\end{equation}
where
\begin{equation}
   D^B_\mu \equiv \partial_\mu - igB_\mu\ .
\end{equation}
This theory should produce the desired magnetic condensate to confine electric
charge.
The magnetic current appearing in (\ref{mgauss4}) is
\begin{equation}
j_g^{\mu}=2g\,{\rm Im}\,\psi^*D^{\mu B}\psi\ .
\end{equation}

We envision a long flux tube with some charge distribution $\pm Q(r)$ at
the ends.
Far from the ends, the flux tube is initially the well-known cylindrically
symmetric solution of the field equations above (and we impose cylindrical
symmetry on the subsequent evolution).
We release into this flux tube a fluid of electrically charged particles,
realized via simple two-fluid MHD.\footnote{This 
means no Schwinger pair creation as yet; it absolves us, however,
of the need to calculate the electric vector potential $A_\mu$.}
The two fluids ($+$ and $-$) obey the Euler equations
\begin{displaymath}
   m \left[\frac{\partial \bv^{\pm}}{\partial t}
      + (\bv^{\pm} \cdot \nabla) \bv^{\pm}\right] = \pm e \bE \pm 
      e \bv^{\pm} \times \bH - \frac{1}{n_e}{\nabla P}
\end{displaymath}
and the continuity equation
\begin{displaymath}
   \frac{\partial n_e}{\partial t} 
      + \nabla \cdot (n_e \bv^{\pm}) = 0\ .
\end{displaymath}
The electric current is thus
\begin{displaymath}
   \bj_e = n_e e \left( \bv^+ - \bv^- \right)\ .
\end{displaymath}
This fluid will flow under the influence of the initial electric field and
will screen the charges at the ends; the fluid's inertia will set up plasma
oscillations.
Since the flux tube geometry is dynamic, the tube itself will react to
the weakening of the field by contracting under the pressure of the
vacuum field $|\psi(\infty)|=v$; then it will be forced open as the fluid
overshoots and oscillates.

The final ingredient is the equation of state of the fluid, 
which we take to be that of a quark--gluon plasma.

\section{PLASMA OSCILLATIONS}

The superconductor has two length scales, the vector mass $m_V=\sqrt2gv$
and the scalar (Higgs) mass $m_H=\sqrt{2\lambda}v$.
In condensed matter language, we have the London penetration depth 
$\lambda_L=m_V^{-1}$ and the coherence length $\xi=m_H^{-1}$; their ratio 
$\kappa=\lambda_L/\xi$ determines whether we have a Type I ($\kappa<1$) or a 
Type II ($\kappa>1$) superconductor.
Another scale in our problem, introduced by the charged fluid, is the
plasma frequency $\omega_p=\sqrt{2n_ee^2/m}$.
If $\omega_p<m_V$, one expects that electromagnetic radiation will be unable
to penetrate into the superconductor, with the Meissner effect providing
dynamical confinement as well as static; the regime $\omega_p>m_V$, where
photons (i.e., Abelian gluons) propagate freely is
presumably outside the range of applicability of the model.

\begin{figure}[htb]
   \begin{center} 
\vskip -10mm
   \mbox{\epsfig{file=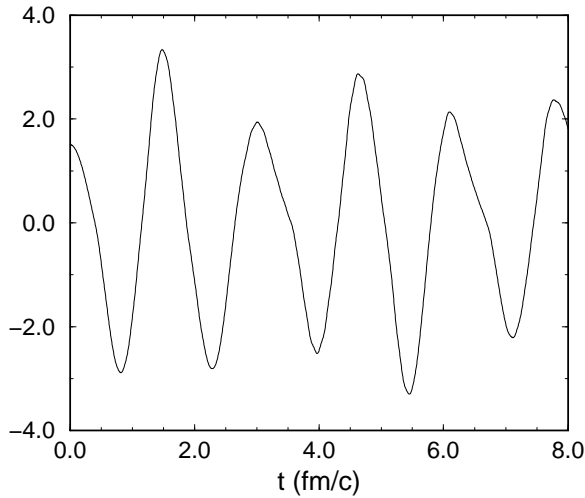,width=3.5in}}
   \end{center}
\vskip -1.7cm
\caption{Plasma oscillations: the electric field $E(0,t)$ on the axis of the flux tube.}
\label{fig:largenenough}
\end{figure}
Our numerical results, presented in the figures, belie our expectations.
We show in Figure 1 the on-axis electric field for the Type I case, with
$\omega_p<m_V$.
There is strong non-linear modification of the plasma oscillations.
Figure 2 presents snapshots of the electric field $E(r)$ and the
Higgs field $\rho(r)=|\psi(r)|$.
It is clear that electric flux penetrates far outside the initial radius
of the flux tube, accompanied by strong oscillations in $\rho$.
(The Type II case is not too different, though less spectacular.)
This surely casts doubt on the usefulness of this model for the study of
dynamical confinement phenomena.
\begin{figure}[htb]
   \begin{center} 
\vskip -1cm
   \mbox{\epsfig{file=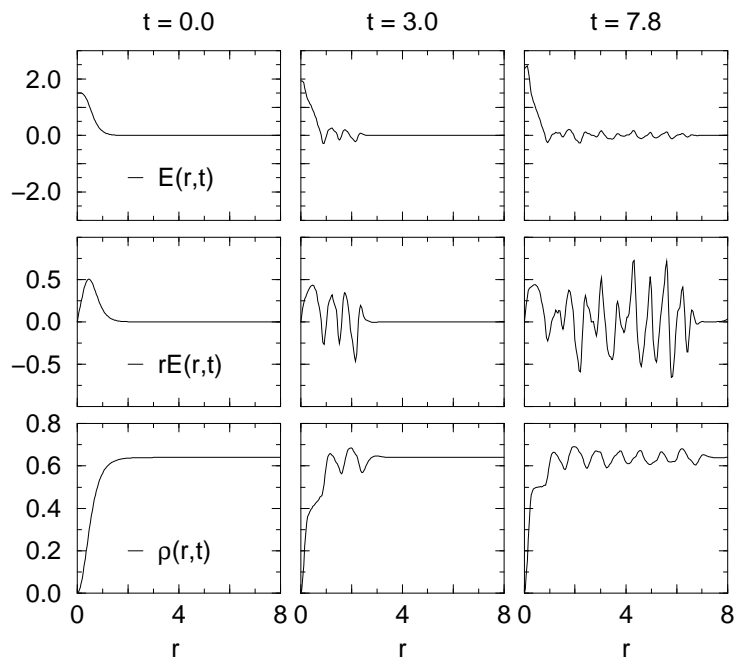,width=5.0in}}
   \end{center}
\vskip -2.3cm
\caption{The electric field $E$ and scalar monopole field $\rho$ at times
corresponding to maxima in the oscillations shown in Figure 1.}
\label{fig:toosmall}
\end{figure}

\end{document}